# Citation Analysis with Medical Subject Headings (MeSH) using the Web of Knowledge: A new routine

Loet Leydesdorff [a] and Tobias Opthof [b,c]

**Abstract**

Citation analysis of documents retrieved from the Medline database (at the Web of Knowledge) has been possible only on a case-by-case basis. A technique is here developed for citation analysis in batch mode using both Medical Subject Headings (MeSH) at the Web of Knowledge and the Science Citation Index at the Web of Science. This freeware routine is applied to the case of "Brugada Syndrome," a specific disease and field of research (since 1992). The journals containing these publications, for example, are attributed to Web-of-Science Categories other than "Cardiac and Cardiovascular Systems"), perhaps because of the possibility of genetic testing for this syndrome in the clinic. With this routine, all the instruments available for citation analysis can now be used on the basis of MeSH terms. Other options for crossing between Medline, WoS, and Scopus are also reviewed.

[a] Amsterdam School of Communication Research (ASCoR), University of Amsterdam, Kloveniersburgwal 48, 1012 CX Amsterdam, The Netherlands; loet@leydesdorff.net; http://www.leydesdorff.net
[b] Experimental Cardiology Group, Heart Failure Research Center, Academic Medical Center AMC, Meibergdreef 9, 1105 AZ Amsterdam, The Netherlands.
[c] Department of Medical Physiology, University Medical Center Utrecht, Utrecht, The Netherlands.

**Introduction**

Comparing different search strategies in the Web of Science, Lundberg *et al.* (2006) concluded that "a well-developed MeSH-based strategy seems to be the preferred method of choice" compared with journals, author names, title words, etc. However, the Medical Subject Headings (MeSH) are available for only part of the journal set contained in multidisciplinary databases such as Scopus or the Web-of-Science (WoS). Other specialist databases such as Chemical Abstracts can equally be considered for improving the precision of the recall because of professional investments in maintaining these indexes (Bensman & Leydesdorff, 2009; Bornmann *et al.*, 2011; cf. Hicks & Wang, 2011).

The WoS and Scopus have been constructed on the basis of selecting journals; journal categories have been added and have periodically been reworked. The MeSH terms, however, are attributed on a paper-by-paper basis to a smaller set of (5500+) journals by the U.S. National Library of Medicine (NLM). Thomson-Reuters (TR)—the current owner of the *Science Citation Index*—adds so-called "topical terms" at the level of papers, but the intellectual basis of these terms is not transparent because both human indexing and computer routines are used. In Scopus, one can search with MeSH terms using a search string within "Indexterms()" (Lutz Bornmann, *personal communication*, June 11, 2012), but the recall is across the database and therefore potentially larger than the recall in PubMed of NLM (at http://www.ncbi.nlm.nih.gov/pubmed/advanced).

Attribution of keywords at the paper level and investments in the thesaurus can be expected to improve on the quality of the delineation and hence the precision of the recall. This is particularly urgent in the case of interdisciplinary developments across journals and journal categories. Hitherto, the WoS Categories (WC) for journals—previously also named the ISI Subject Categories—have been used for the evaluation of (interdisciplinary) research and mapping (e.g., Rafols *et al.*, 2010). However, Boyack *et al.* (2005) and Pudovkin & Garfield (2002, at p. 1113n.) already expressed concerns about the quality of the attributions (Rafols & Leydesdorff, 2009). Bornmann *et al.* (2008) suggested that MeSH terms or other specialized index terms might provide reference sets for the evaluation that would be more precise than journal sets.

In addition to the problem of delineating groups of journals, journals may themselves be heterogeneous in intellectual terms (Boyack & Klavans, 2011), and researchers may deliberately cross boundaries in interdisciplinary research (Rafols *et al.*, 2012). However, despite the availability of Medline and MeSH terms in the Web-of-Knowledge (WoK) juxtaposed to the *Science Citation Index* in the Web-of-Science (WoS), this integration of these two databases could hitherto not be used for citation analysis. Version 5 of WoS, available since August 2011, made citation rates (from WoS) visible in the results of Medline searches in WoK (insofar as the journals are part of the overlap). However, one has to enter each record one-by-one to find these values, and the citation scores are not included in the download of document sets from this environment.

In this brief communication, we report on how to exploit additionally the information of the WoS accession number in the Medline database for the construction of an interface that makes citation analysis possible using MeSH terms. The freeware routine is part of a series of studies in which (in collaboration with different co-authors; e.g., Bornmann & Leydesdorff, 2011; Leydesdorff *et al*., in preparation; Leydesdorff & Bornmann, in press; Leydesdorff & Persson, 2010; Rafols *et al*., 2010) we aim to integrate datasets using common baselines such as Google Maps or indexes such as the *Index Medicus* in order to develop multiple perspectives on the same (or similar) data. Multiple perspectives may enable a user to appreciate the non-linear dynamics of innovations as trajectories of clusters in multivariate spaces (that can be mapped using multidimensional scaling, etc.). Breaking the barrier between the Medline database and WoS can be considered as an important step from the perspective of this longer-term research program.

**Methods and materials**

The argument will be developed using the MeSH term "Brugada Syndrome" (BrS) as an example. BrS is a rare cardiac disease which may lead to fatal arrhythmias and thereby to sudden death (Sudden Unexpected Death Syndrome; SUDS). The incidence of BrS is higher in males, and both genetic and geographical aspects are involved, since the disease is familiar and more common in Southeast Asia. The debate over the cause(s) of BrS is far from settled.[1] BrS leads to aberrations in the electrocardiogram (ECG) with mild voltage elevation at the transition of the QRS complex into the ST segment of the ECG (Brugada & Brugada, 1992).

The community of researchers focusing on this issue is relatively small, and the publication output is accordingly in the order of 100-200 publications per year. We selected the topic because it is very specific and we have access to the expertise of bio-medical scientists working in this community. BrS has been indexed in Medline/PubMed since 2007 under the MeSH terms C14.280.067.322 as a cardiac arrhythmia and C16.320.100 as a congenital disease. We searched with the string '"Brugada syndrome"[MeSH Terms] AND ("2010.01.01"[PDAT] : "2011.31.12"[PDAT]).' This provided a recall of 287 records on February 24, 2012. A similar search recalled 286 (= 287 − 1) records in the Medline installation of Thomson Reuters at WoK.

---

[1] The debate is about whether the disease should be considered a 'conduction' or a 'repolarization' disease (Wilde *et al*., 2010). The hearts of BrS patients appear to have structural abnormalities (fibrosis) at autopsy that remain clinically unnoticed (Hoogendijk *et al*., 2010). Furthermore, there is debate about the value of programmed electrical stimulation by which the vulnerability to arrhythmias is tested (Priori *et al*., 2012) with the aim of indicating patients at risk of sudden death. Positive outcomes (false ones and true ones) have led to the implantation of internal cardiac defibrillators. The majority of these implanted devices cause trouble although occasionally saving lives.

Two dedicated software routines were developed: the first one, called Medline.exe, retrieves the data at the Internet using the so-called MicroSoft Internet Protocol. The user may have to install this protocol, because not all versions of Windows install the protocol initially.[2] The Medline data from WoK is saved on disk as html using sequential numbering as follows: p1.htm, p2.htm, etc. The second routine, called MH2WoS.exe organizes this data into relational databases, and produces a file "wos.txt" which can be used for retrieval at the Advanced Search interface of WoS. The field "UT" for the accession number of WoS is used for this retrieval. If the search string is too long, the user can edit it. The "times cited" are additionally stored in the file TI.dbf which contains the output of MH2WoS.exe. However, this file includes also the Medline records that are not included in WoS (and thus cannot be rated in terms of citations using WoS). TI.dbf also contains the PubMed Identification Numbers (PMID) which can be used for searching in Scopus, but again the coverage may not be complete.

The routine MH2WoS.exe calls Medline.exe optionally (if one has not first downloaded the data). Medline.exe first prompts the user for the search string of *any* one of the records retrieved within the WoK set. Thus: the user has to enter an arbitrary record among the results—which one of the set is not important because the string is parsed by the routine—and then to copy (Ctrl-C) the string from the navigation bar of the browser. This string contains all information needed for the analysis. The user is further prompted for the starting number in the set (default is 1) and the number of documents to be retrieved. The download may take a bit of time; note that savings from previous downloads with the same name are overwritten. Both the routines and the files are to be saved in a single folder.

After finishing the retrieval at the Internet, the routine MH2WoS.exe organizes this data into three (relational) databases: AU.dbf, MH.dbf, and TI.dbf. TI.dbf contains the information that is unique to each document, including the "times cited" (TC) at the moment of the download. MH.dbf contains the MeSH terms and—if present—the so-called qualifiers of MeSH. The address information is not further parsed because this information can be made more complete and standardized using the WoS in the possible follow-up step. To obtain the documents in WoS, one can use the file "wos.txt" as input to the Advanced Search option of WoS. (In the case of a large number, one may wish to edit this file and combine the recalls within WoS.) The routines MH2WoS.exe and Medline.exe can be retrieved at http://www.leydesdorff.net/medline. This webpage also contains further (and potentially updated) instructions.

**"Brugada Syndrome"**

Of the 286 records retrieved from Medline (at WoK) with the MeSH "Brugada syndrome" and publication years in 2010 or 2011, only 235 contained a WoS identifier; 114 of these records were cited 435 times from publication till February 24, 2012, whereas 121 remained uncited at that time. Figure 1 shows a new download of citations at June 26, 2012 with comparable, but of course slightly other numbers (see legend Figure 1). The output file wos.txt (of MH2WOS.exe) contains the 235 identifiers concatenated as follows: "UT= (000298415800028 OR 000297149900006 OR …)".

---

[2] See for installation instructions at http://www.leydesdorff.net/software/patentmaps/ocx.htm (cf. Leydesdorff & Bornmann, in press). On a network installation, one may not have administrative rights for installing this protocol.

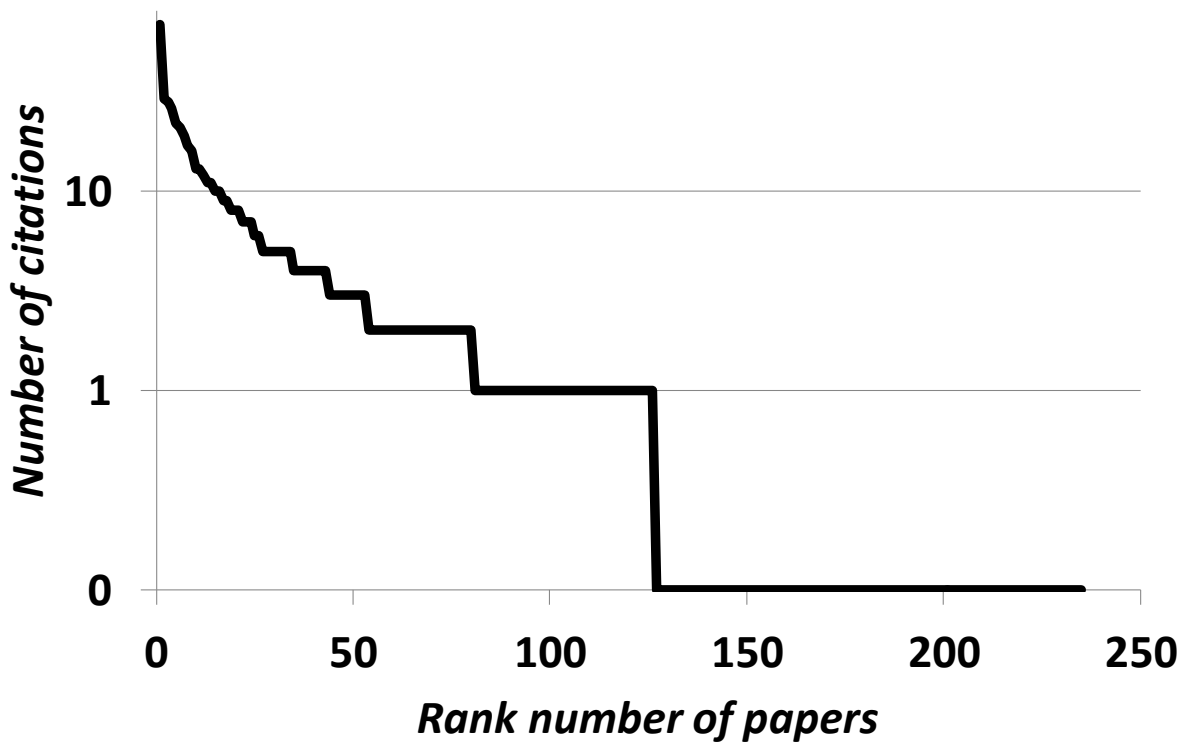

**Figure 1**: Ranking of 608 citations to 235 papers with MeSH category "Brugada syndrome" and published in 2010 or 2011. Retrieved on June 26, 2012; (235 − 126 =) 109 papers in the set were not cited until this date.[3]

How homogeneous is this set when compared with the Web-of-Science categories (WC)? the distribution of the 292 WCs attributed to these 235 papers is as follows (Table 1):

| Web of Science Categories | $N$ |
|---|---|
| Cardiac Cardiovascular Systems | 186 |
| Medicine General Internal | 24 |
| Engineering Biomedical | 10 |
| Peripheral Vascular Disease | 10 |
| Hematology | 8 |
| Physiology | 8 |

---

[3] The citation analysis was repeated on June 26, 2012 after referee comments. At this date, 319 records could be retrieved at PubMed of which 316 could also be found at the Medline interface of WoK.

| | |
|---|---|
| Emergency Medicine | 7 |
| Clinical Neurology | 4 |
| Critical Care Medicine | 4 |
| Pharmacology Pharmacy | 4 |
| Anesthesiology | 3 |
| Pediatrics | 3 |
| Public Environmental Occupational Health | 3 |
| Sport Sciences | 3 |
| Biochemical Research Methods | 2 |
| Cell Biology | 2 |
| Chemistry Analytical | 2 |
| Genetics Heredity | 2 |
| Surgery | 2 |
| Biology | 1 |
| Medicine Legal | 1 |
| Neurosciences | 1 |
| Nursing | 1 |
| Obstetrics Gynecology | 1 |

**Table 1**: Distribution of WoS Categories attributed to 235 documents from a highly specialized field ("Brugada Syndrome").

The attribution of WoS Categories in neighboring fields in addition to the main category of "Cardiac and Cardiovascular Systems" is not surprising (Rafols *et al.*, 2010), since journals can be attributed to more than a single WC. More telling, in our opinion, is that only 186 of these 235 papers (79%) were attributed to the core category.

For example, one of the papers entitled "Decadal Electrocardiographic Changes Between Age 40 and 50 in Military Pilots" (Ohrui *et al.*, 2011) is attributed to the WCs (*i*) Public, Environmental & Occupational Health, (*ii*) Medicine, General & Internal, and (*iii*) Sport Sciences, but not to the WC for cardiac systems although the Brugada syndrome is mentioned in the abstract of this paper as one of the possible causes of sudden death. As noted, attribution of WCs occurs not at the level of articles, but at the level of journals. The journal in this case ("*Aviation Space and Environmental Medicine*") is not attributed to the category of "Cardiac & Cardiovascular Systems." In sum, the indexing at the paper level by NLM provides us with more specific sets than the journal indexing at TR.

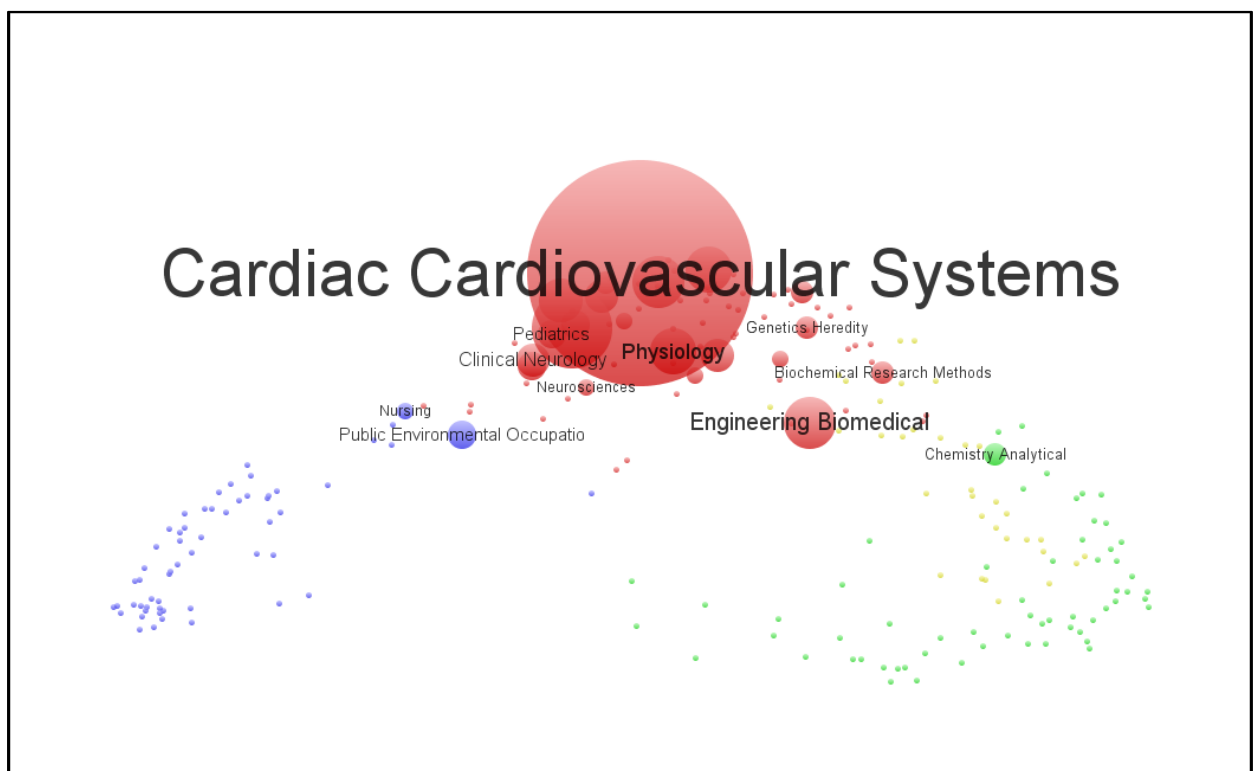

**Figure 2**: Attribution of the set of 235 papers classified with the MeSH "Brugada Syndrome" in 2010 or 2011 to 24 WCs. Rao-Stirling diversity = 0.3684. VOSViewer v. 1.4.1 used for the mapping.

In order to show how neatly the output of this new instrument can work with the set of tools which we made previously available for mapping WoS data both geographically and intellectually, Figure 2 shows the map of these WCs using VOSviewer for the visualization (Rafols *et al.*, 2010; Leydesdorff, Carley, & Rafols, in preparation). The Rao-Stirling diversity (which can vary between zero and one) is equal to 0.37, and thus a certain interdisciplinarity is indicated (Leydesdorff & Rafols, 2011; Rafols& Meyer, 2010; Rafols *et al.*, 2010; Stirling, 2007).

These results suggest that "Brugada Syndrome" is no longer only a highly specialized subject of research in the laboratory, but has in the meantime been diffused into medical practices (Leydesdorff, Rotolo & Rafols, 2012). A genetic test for Brugada Syndrome is nowadays clinically available. This also makes it possible to determine if relatives of victims are at risk of the disease.

**Conclusions and discussion**

The new tool (MH2WoS.exe) enables the user to perform citation analysis in WoS using MeSH terms after a step of pre-processing in WoK. As noted, citation analysis with PubMed identifiers (PMIDs) is available within Scopus. These identifiers are also stored (and therefore available) in the file TI.dbf for the user, but cannot be used for searching at WoS. For this purpose one needs the WoS identifiers.

The reverse road of using WoS search results in the PubMed database (at http://www.ncbi.nlm.nih.gov/pubmed/advanced) requires an in-between step at http://www.ncbi.nlm.nih.gov/pubmed/batchcitmatch for the identification of PMIDs. (This step is further explained in Leydesdorff, Rotolo, & Rafols (in press, section 7.2).)

As noted, Scopus output already contains these PMIDs. One limitation of the current routines is that documents that are included in Medline/PubMed, but not in WoS or Scopus, cannot be analyzed in terms of citations. The problem that the Medline database provides unsystematic address information can be circumvented, and Google Map overlays can be generated using routines provided previously (Bornmann & Leydesdorff, 2011; Leydesdorff & Persson, 2010).

In summary, a missing link in the relations between PubMed, WoS, and Scopus was thus constructed. Let us summarize the possibilities and relationships as follows:

1. Using PubMed, one can search and retrieve using MeSH terms or PMID, and a set of other qualifiers, but one cannot retrieve citation rates. Output of Scopus contains PMID, but output from WoS requires an in-between step at http://www.ncbi.nlm.nih.gov/pubmed/batchcitmatch for the identification of PMIDs before it can be used as input to PubMed;
2. Using Scopus, one can search with PMID (obtained, for example, in PubMed) or directly with MeSH terms. However, the recall with MeSH terms includes papers which are not included in PubMed because the thesaurus of NLM is used in this differently defined database. For example, repeating the analysis with the same search string on June 26, 2012 provided us with a recall of 319 papers in PubMed, 315 in Medline at WoK, but 580 at Sopus using the searchstring 'Indexterms("Brugada syndrome" AND ((pubyear is 2010) or (pubyear is 2011))'. Below 2000 records, Scopus allows for downloading and thus one can further identify whether or not each record has the PMID(s) under study;
3. Using WoS/WoK and our routines, one can integrate MeSH searches with citation analysis for sets sized below the systems limitation of 100,000 in WoS.

**Acknowledgement**
The authors are grateful to Lutz Bornmann, Ismael Rafols, and Daniele Rotolo for input and feedback to this study.